\documentclass[runningheads]{llncs}

\usepackage[T1]{fontenc}
\usepackage{graphicx}
\usepackage{color}
\usepackage{algorithm}
\usepackage{comment}
\usepackage[noend]{algpseudocode}
\usepackage{amsmath}
\usepackage{hyperref}
\usepackage{enumitem}
\usepackage{booktabs}
\usepackage{multirow}

\usepackage[table,xcdraw]{xcolor}
\usepackage[table]{xcolor}

\algnewcommand{\LineComment}[1]{\State \(\triangleright\) #1}

\definecolor{lightgray}{gray}{0.9}

\begin{document}

\title{Grid-Based Decompositions for Spatial Data under Local Differential Privacy}

\titlerunning{Grid-Based Decompositions Under LDP}
%

\author{Berkay Kemal Balioglu \and Alireza Khodaie \and
Ameer Taweel \and \\ M. Emre Gursoy}
\authorrunning{B. K. Balioglu et al.}

\institute{Department of Computer Engineering, Koç University, Istanbul, Turkey\\
\email{\{bbalioglu23, akhodaie22, ataweel20, emregursoy\}@ku.edu.tr}}
\maketitle              

\begin{abstract}
Local differential privacy (LDP) has recently emerged as a popular privacy standard. With the growing popularity of LDP, several recent works have applied LDP to spatial data, and grid-based decompositions have been a common building block in the collection and analysis of spatial data under DP and LDP. In this paper, we study three grid-based decomposition methods for spatial data under LDP: Uniform Grid (UG), PrivAG, and AAG. UG is a static approach that consists of equal-sized cells. To enable data-dependent decomposition, PrivAG was proposed by Yang et al.~as the most recent adaptive grid method. To advance the state-of-the-art in adaptive grids, in this paper we propose the Advanced Adaptive Grid (AAG) method. For each grid cell, following the intuition that the cell's intra-cell density distribution will mimic its neighbors, AAG performs uneven cell divisions depending on the neighboring cells' densities. We experimentally compare UG, PrivAG, and AAG using three real-world location datasets, varying privacy budgets, and query sizes. Results show that AAG provides higher utility than PrivAG, demonstrating the superiority of our proposed approach. Furthermore, UG's performance is heavily dependent on the choice of grid size. When the grid size is chosen optimally in UG, AAG still beats UG for small queries, but UG beats AAG for large (coarse-grained) queries.


\end{abstract}

\section{Introduction} \label{sec:Introduction}
\vspace{-2pt}

Large volumes of spatial data are nowadays available for collection and analysis, thanks to the popularity of smartphones, connected cars, location-based services (LBS), and social networks. There is great value in collecting and analyzing this data for deriving insights and improving products and services. On the other hand, ensuring the privacy of spatial data is imperative since it contains sensitive information about individuals, such as their home and work addresses, frequently visited locations, and personal habits. In recent years, differential privacy has become a widely accepted standard for privacy protection. Differential privacy has two main flavors: centralized DP which assumes the existence of a trusted aggregator, and local DP (LDP) which removes the need for a trusted aggregator. In LDP, each user locally perturbs their data on their own device and shares the perturbed version with the aggregator. LDP has recently received significant attention from the academia and industry, and it has been deployed in products of various tech companies such as Apple, Google, and Microsoft \cite{cormode2018privacy,ding2017collecting,erlingsson2014rappor,gursoy2022adversarial}. 

With the growing popularity of LDP, several recent works have applied LDP to spatial data and LBS \cite{alptekin2023building,du2023ldptrace,hong2022collecting,wang2021lsrr,yang2022collecting,zhang2023trajectory}. However, in many applications of DP and LDP to spatial data, the data needs to be discretized so that the input and output domains of the privacy mechanisms are discrete and finite. Grid-based decompositions, which decompose the overall geospatial area $\Omega$ into non-overlapping cells, are a popular method for this purpose. After a grid is laid, the user's location can be discretized by determining which cell it falls inside. Indeed, uniform and adaptive grids have been widely used in the DP and LDP literature for trajectory collection and sharing \cite{du2023ldptrace,yang2022collecting}, range query answering \cite{qardaji2013differentially}, synthetic data generation \cite{gursoy2018differentially,gursoy2018utility}, and so forth.

A uniform grid (UG) partitions the geospatial area $\Omega$ into $N \times M$ cells of equal size. However, the uniform grid is a static approach that does not adapt to the underlying data distribution \cite{qardaji2013differentially,yang2022collecting}. It may result in a poor partitioning of $\Omega$ when certain regions of $\Omega$ are too crowded or too sparse (i.e., too high density or too low density). To enable density-dependent decomposition of $\Omega$, the adaptive grid approach was proposed \cite{qardaji2013differentially,yang2022collecting,gursoy2018utility}. The main idea of an adaptive grid is to first lay a uniform grid $\mathcal{G}_1$ over the given $\Omega$, and according to the cell density estimations obtained from users, further divide each individual cell (i.e., adapt the grid). To the best of our knowledge, the most recent adaptive grid approach in LDP is PrivAG proposed by Yang et al.~\cite{yang2022collecting}.

In this paper, we advance the state-of-the-art in adaptive grids by proposing Advanced Adaptive Grid (AAG). In PrivAG, when a certain cell $C_k \in \mathcal{G}_1$ needs to be divided further, this division is done evenly. In AAG, we propose to perform this division by taking into account $C_k$'s neighboring cells. This is because $C_k$'s intra-cell density distribution is likely to be affected by its neighbors. For example, if $C_k$'s right neighbor is dense but the left neighbor is sparse, then the intra-cell density of $C_k$ is likely to be skewed towards the right. Following the intuition behind adaptive grids which is to place many small cells in dense areas but few large cells in sparse areas, instead of dividing $C_k$ evenly, we perform an uneven division in AAG which is weighted proportional to the neighbors' densities. Furthermore, we propose heuristic strategies to handle edge cells and corner cells which lack one or more neighbors. In addition, motivated by our empirical observation that the parameter choices in PrivAG yield adaptive grids with cell counts that are too similar to $\mathcal{G}_1$, we propose new parameter values for AAG.

We experimentally compare the three grid approaches (UG, PrivAG, AAG) using three real-world location datasets (Gowalla, Porto, Foursquare) by measuring their errors in answering spatial density queries. We measure their Average Query Errors (AQE) with different privacy budgets $\varepsilon$ and query sizes $\rho$. We find that the AQEs of UG are heavily dependent on the grid size, i.e., it performs well when the grid size is chosen optimally, but it can perform quite poorly otherwise. After determining optimal grid sizes for UG and then comparing it with PrivAG and AAG, we observe that: (i) AAG is preferable to PrivAG across all $\varepsilon$ and $\rho$, demonstrating that our AAG improves the state-of-the-art adaptive grid approach in LDP, (ii) AAG is the best approach when $\rho$ is small, and (iii) UG is the best approach when $\rho$ is large, e.g., for computing answers to large queries or for fetching coarse (high-level) statistics.

\textbf{Contributions.} In summary, our main contributions are the following:
\vspace{-3pt}
\begin{itemize}
    \item We propose a novel approach called Advanced Adaptive Grid (AAG) for grid-based spatial data decompositions under LDP. 
    \item We experimentally show that AAG yields lower error compared to the state-of-the-art adaptive grid approach in LDP, called PrivAG \cite{yang2022collecting}.
    \item We show that the errors of UG are highly dependent on the grid size. When the grid size is chosen optimally, then AAG still beats UG for small queries. Yet, UG is the best approach when the queries are large. 
\end{itemize}


\vspace{-12pt}
\section{Background} \label{sec:Background}

\vspace{-4pt}
\subsection{Data Model and Notations} \label{sec:DataModel}

Let $\mathcal{U} = \{u_1, u_2, u_3, ...\}$ denote the set of users in the system where $|\mathcal{U}|$ is the total number of users, and let the two-dimensional geospatial area be denoted by $\Omega$. For each user $u_i$, the user's true location is represented by $l_i$, such that $l_i \in \Omega$ and $l_i$ consists of a pair of \textit{(latitude, longitude)} coordinates. We assume that the boundaries of the overall domain $\Omega$ are not privacy-sensitive, and therefore, they can be known by all parties in the system. On the other hand, each user's location is privacy-sensitive and must be protected. For example, if $\Omega$ corresponds to the city of Istanbul, the boundaries of Istanbul are publicly known. In contrast, the locations of each user within Istanbul need to be protected. 


\vspace{-4.5pt}
\subsection{Local Differential Privacy} \label{sec:LDP}

Local differential privacy (LDP) is a widely accepted standard for safeguarding privacy, and it has been used in various products of tech companies such as Apple, Google, and Microsoft \cite{ding2017collecting,erlingsson2014rappor}. In LDP, users' data is perturbed on their devices before being collected by the aggregator (also called the ``server''). After collecting perturbed data from users, the server uses estimation methods to recover statistics pertaining to the general population $\mathcal{U}$. However, since each user's data is perturbed, the server cannot infer exact information about a specific user. In our context, since the private information that needs to be protected by LDP is each user's location $l_i$, we define LDP as follows. 

\begin{definition}[$\varepsilon$-LDP] \label{def:LDP}
\textnormal{A randomized algorithm $\Psi$ satisfies $\varepsilon$-local differential privacy ($\varepsilon$-LDP), where $\varepsilon > 0$, if and only if for any two inputs $l_i,l^*_i$, it holds that:
\begin{equation}
\forall y \in Range(\Psi): ~~~~ \frac{{Pr}[\Psi(l_i) = y]}{{Pr}[\Psi(l^*_i) = y]} \leq e^{\varepsilon}
\end{equation}
where $Range(\Psi)$ stands for the set of all possible outputs of the algorithm $\Psi$.}
\end{definition}

$\varepsilon$-LDP ensures that having observed the output $y$, the server (or any other party who observed $y$) is not able to distinguish whether the original location of the user was $l_i$ or $l^*_i$ with probability more than the odds ratio controlled by $e^\varepsilon$. The strength of the privacy protection is controlled by the parameter $\varepsilon$, commonly known as the \textit{privacy budget}. Lower $\varepsilon$ yields stronger privacy.

\vspace{-3pt}
\subsection{Optimized Local Hashing} \label{sec:OLH}
\vspace{-1pt}

Numerous LDP protocols have been proposed to minimize utility loss and/or communication cost under various conditions. In this paper, we use a state-of-the-art protocol called \textit{Optimized Local Hashing (OLH)} \cite{wang2017locally} due to its high utility and low communication cost \cite{cormode2021frequency,gursoy2022adversarial}. Similar to other LDP protocols, OLH consists of two main components: (i) user-side encoding and perturbation to satisfy LDP on users' devices, and (ii) server-side estimation after collecting perturbed data from the user population.

\textbf{User-Side Perturbation in OLH:} OLH uses hash functions to reduce the size of the input domain $\mathcal{D}$. To address the issue of recurring hash collisions, a family of hash functions $\mathcal{H}$ is used such that each user chooses a different hash function from $\mathcal{H}$. Let $\mathcal{H}$ be a hash function family such that each $H \in \mathcal{H}$ maps a value from $\mathcal{D}$ to an integer between 1 and $m$, i.e., $H: \mathcal{D} \rightarrow [1,m]$. User $u$ with true value $v_u$ chooses a hash function from $\mathcal{H}$ uniformly at random $H_u \leftarrow_{\$} \mathcal{H}$ and computes the integer $x_u \gets H_u(v_u)$. Then, the perturbation step takes $x_u$ as input and produces a perturbed integer $x'_u \in [1,m]$ with the probabilities:
\begin{equation} \label{eq:OLH}
\forall_{i \in [1,m]} :~ \text{Pr}[x'_u=i] = 
\begin{cases}
\frac{e^{\varepsilon}}{e^{\varepsilon}+m-1} & \text{ if } x_u=i \\
\frac{1}{e^{\varepsilon}+m-1} & \text{ if } x_u \neq i
\end{cases}
\end{equation}
The user sends the resulting tuple $\langle H_u,x'_u \rangle$ to the server. Following \cite{wang2017locally,wang2018locally}, we use the default value of $m$ as $m=e^\varepsilon+1$, which optimizes utility.  

\textbf{Server-Side Estimation in OLH:} The server receives tuples from users $u \in \mathcal{U}$. To perform estimation for some value $v \in \mathcal{D}$, the server first computes $Sup(v)$ = total number of tuples for which the condition $x'_u = H_u(v)$ holds. Then, the server estimates the number of occurrences of value $v$ in the population using the following unbiased estimator $\Phi(v)$:
\begin{equation} \label{eq:aggregator-OLH}
\Phi(v) = \frac{(e^\varepsilon+m-1) \cdot (m \cdot Sup(v) - |\mathcal{U}|)}{(e^\varepsilon-1) \cdot (m-1)}
\end{equation}

\section{Grid-Based Decompositions Under LDP} \label{sec:Grids}


Grid-based decompositions are commonly used in the DP and LDP literature for spatial data collection and analysis \cite{du2023ldptrace,gursoy2018differentially,gursoy2018utility,qardaji2013differentially,yang2022collecting}. They employ uniform or adaptive grid structures to partition the geospatial area $\Omega$ into cells. Then, the user's location $l_i$ is discretized by determining which cell it falls inside. We first describe the Uniform Grid (UG) approach in Section \ref{sec:UniformGrid}, then the existing adaptive grid approach called PrivAG \cite{yang2022collecting} in Section \ref{sec:PrivAG}, and finally our novel Advanced Adaptive Grid approach called AAG in Section \ref{sec:AAG}. 

\subsection{Uniform Grid (UG)} \label{sec:UniformGrid}

A uniform grid partitions the geospatial area $\Omega$ into $N \times M$ cells of equal size. We denote this grid by $\mathcal{G}_{uni} = (C_1, C_2, ..., C_{N \times M})$ where each $C_j \in \mathcal{G}$ is one cell. The geographic coverage of all cells are disjoint from one another. User $u_i$ with location $l_i$ discretizes his/her location by finding which $C_i \in \mathcal{G}$ their $l_i$ falls inside. Then, to satisfy LDP, the cell information $C_i$ needs to be perturbed. Hence, we feed $C_i$ into the OLH protocol with the appropriate parameters.  

\begin{algorithm}[!t]
\caption{Collecting location data with LDP using a grid}
\label{alg:uniform}
\begin{algorithmic}[1]
\State \textbf{Input:} Users $\mathcal{U}$, grid $\mathcal{G}$, privacy budget $\varepsilon$
\State \textbf{Output:} Densities of each cell in $\mathcal{G}$
\\
\LineComment{\textbf{User-side discretization and perturbation}}
\For{each user $u_i \in \mathcal{U}$} 
\For{each cell $C_j \in \mathcal{G}$}
    \If{$l_i$ falls inside $C_j$}
        \State Set user's true cell as: $C_i \gets C_j$ 
        \State \textbf{break} 
    \EndIf
\EndFor
\State Execute the user-side OLH protocol with true value = $C_i$, domain $\mathcal{D}$ = $\mathcal{G}$, and privacy budget = $\varepsilon$
\State Send the resulting tuple $\langle H_u,x'_u \rangle$ to the server
\EndFor

\\

\LineComment{\textbf{Server-side estimation}}
\State Server receives $\langle H_u,x'_u \rangle$ from all $u_i \in \mathcal{U}$
\For{each cell $C_j \in \mathcal{G}$}
    \State Compute $Sup(C_j)$ as the number of tuples for which $x'_u = H_u(C_j)$
    \State Compute estimate $\Phi(C_j)$ using Equation \ref{eq:aggregator-OLH}
\EndFor
\State \textbf{return} $\Phi(C_1)$, $\Phi(C_2)$, ... for all $C_j \in \mathcal{G}$

\end{algorithmic}
\end{algorithm}

An overview of LDP location data collection using $\mathcal{G}_{uni}$ is shown in Algorithm \ref{alg:uniform}. For each user $u_i$ with location $l_i$, the user first finds their true cell $C_i$, i.e., which cell in the grid their location falls inside (lines 6-9). After finding $C_i$, the user executes the OLH protocol by treating their true value as $C_i$, the domain of the protocol as $\mathcal{G}_{uni} = (C_1, C_2, ..., C_{N \times M})$, and using the privacy budget $\varepsilon$. Since OLH satisfies $\varepsilon$-LDP, the perturbation of $C_i$ using OLH achieves $\varepsilon$-LDP. Note that since $l_i$ is discretized as $C_i$, the domain of OLH is also discretized: $\mathcal{D} = \mathcal{G}_{uni} = (C_1, C_2, ..., C_{N \times M})$, instead of using a continuous domain $\mathcal{D} = \Omega$. Each user sends the OLH protocol output to the server. The server receives the outputs from all users and then proceeds to estimate the density of each cell $C_j \in \mathcal{G}_{uni}$. Here, the density of a cell corresponds to the number of users whose locations are in that cell. The estimation of each cell $C_j \in \mathcal{G}_{uni}$ is performed one by one (lines 15-17), using the server-side estimation procedures of OLH explained in Section \ref{sec:OLH}. Finally, the estimated densities are returned (line 18).

Note that although Algorithm \ref{alg:uniform} is given with the OLH protocol, it is possible to collect location data using a uniform grid with other LDP protocols such as GRR, RAPPOR, and OUE \cite{gursoy2022adversarial,wang2017locally}. The general intuition and procedure remain the same, but the user-side perturbation (lines 10-11) and server-side estimation (lines 16-17) steps would be changed according to the LDP protocol in use. 

\subsection{Existing Adaptive Grid: PrivAG} \label{sec:PrivAG}


UG is a static approach that does not adapt to the underlying data distribution \cite{qardaji2013differentially,yang2022collecting}. It may result in a poor partitioning of $\Omega$ in scenarios where certain regions are too crowded or too sparse (too high density or too low density). For example, when a cell is too crowded, then further partitioning it into smaller cells enables a better understanding of the detailed data distribution \textit{within} that cell. Yet, the uniform grid is not able to achieve this. On the other hand, if a certain region of $\Omega$ is sparse, then the cells in that region will have zero or near-zero density, and the uniform grid will be over-partitioning that region. Over-partitioned cells with zero density lead to utility loss since their estimated densities are likely to be non-zero due to LDP perturbation, leading to fictitious and skewed results. To overcome these limitations and enable data-dependent decompositions, the adaptive grid approach was proposed in DP and LDP \cite{qardaji2013differentially,yang2022collecting}. The main idea of an adaptive grid is to first lay a uniform grid $\mathcal{G}_1$ over the given $\Omega$, and according to the cell density estimations obtained from users, further divide each individual cell (i.e., adapt the grid) to obtain the final grid. 

To the best of our knowledge, the most recent adaptive grid approach in LDP is PrivAG, proposed by Yang et al.~\cite{yang2022collecting}. In PrivAG, the server first divides the set of users $\mathcal{U}$ into two groups: $\mathcal{U}_1$ and $\mathcal{U}_2$. Then, the server constructs a uniform grid $\mathcal{G}_1$ of size $g_1 \times g_1$ and broadcasts $\mathcal{G}_1$ to users in $\mathcal{U}_1$. Based on $\mathcal{G}_1$, users in $\mathcal{U}_1$ discretize their locations and use OLH to send their perturbed outcomes to the server. The server estimates the densities of each cell in $\mathcal{G}_1$. Then, for each cell $C_k \in \mathcal{G}_1$, the server further partitions $C_k$ into $g_2^k \times g_2^k$ cells, where $g_2^k$ depends on $\Phi(C_k)$ and several other parameters (more detail below). After all cells are partitioned according to their $g_2^k$, the resulting adaptive grid $\mathcal{G}_{ag}$ is obtained. Then, $\mathcal{G}_{ag}$ is advertised to users in $\mathcal{U}_2$ and desired statistics are obtained using $\mathcal{G}_{ag}$, e.g., cell densities or Hidden Markov Model weights \cite{yang2022collecting}. 

\textbf{Variables $g_1$ and $g_2^k$.} Since the values of $g_1$ and $g_2^k$ have an important impact on the grid that is constructed, Yang et al.~\cite{yang2022collecting} propose guidelines for choosing them. According to their guidelines, $g_1$ should be chosen as:
\begin{equation} \label{g1_calc}
    g_1 = \sqrt{2 \alpha \cdot (e^{\varepsilon} - 1) \cdot \sqrt{\frac{|\mathcal{U}|}{e^{\varepsilon}}}}    
\end{equation}
Then, for each cell $C_k$ with estimated density $\Phi(C_k)$, its $g_2^k$ should be chosen as:
\begin{equation} \label{g2_calc}
    g_2^k = \sqrt{2 \alpha \cdot \Phi(C_k) \cdot (e^{\varepsilon} - 1) \cdot \sqrt{\frac{(1 - \sigma) \cdot |\mathcal{U}|}{e^{\varepsilon}}}}
\end{equation}

Here, $\alpha$ is a small constant recommended to take values between [0.01, 0.02], and $\sigma$ denotes the proportion of users in $\mathcal{U}_1$, i.e., $\sigma = \frac{|\mathcal{U}_1|}{|\mathcal{U}|}$. It is recommended for $\sigma$ to be in the range [0.1, 0.3]. An algorithmic overview of the PrivAG approach is given in Algorithm \ref{alg:PrivAG}. 

\begin{algorithm}[!t]
\caption{Algorithmic summary of the PrivAG approach}
\label{alg:PrivAG}
\begin{algorithmic}[1]
\State \textbf{Input:} Users $\mathcal{U}$, parameters $\alpha$ and $\sigma$, privacy budget $\varepsilon$
\State \textbf{Output:} Densities of each cell in adaptive grid $\mathcal{G}_{ag}$
\\
\LineComment{\textbf{First phase of PrivAG}}
\State Server computes $g_1$ according to Equation \ref{g1_calc}
\State Server divides $\mathcal{U}$ into two groups $\mathcal{U}_1$ and $\mathcal{U}_2$ such that $|\mathcal{U}_1|$ = $\sigma \times |\mathcal{U}|$
\State Server lays $g_1 \times g_1$ uniform grid $\mathcal{G}_1$ on $\Omega$
\State Call Algorithm \ref{alg:uniform} with $\mathcal{U}_1$, $\mathcal{G}_1$ and $\varepsilon$ to obtain $\Phi(C_1)$, $\Phi(C_2)$, ... for all $C_k \in \mathcal{G}_1$

\\
\LineComment{\textbf{Second phase of PrivAG}}
\For{each cell $C_k \in \mathcal{G}_1$}
    \State Compute $g_2^k$ according to Equation \ref{g2_calc}
    \State Divide $C_k$ into $g_2^k \times g_2^k$ cells of equal size
\EndFor
\State Let $\mathcal{G}_{ag}$ denote the resulting grid after the above divisions
\State Call Algorithm \ref{alg:uniform} with $\mathcal{U}_2$, $\mathcal{G}_{ag}$ and $\varepsilon$ to obtain $\Phi(C_1)$, $\Phi(C_2)$, ... for all cells in $\mathcal{G}_{ag}$

\end{algorithmic}
\end{algorithm}

\vspace{-4pt}
\subsection{Proposed Approach: Advanced Adaptive Grid (AAG)} \label{sec:AAG}

\begin{figure}[!t]
    \centering
    \includegraphics[width=.75\textwidth]{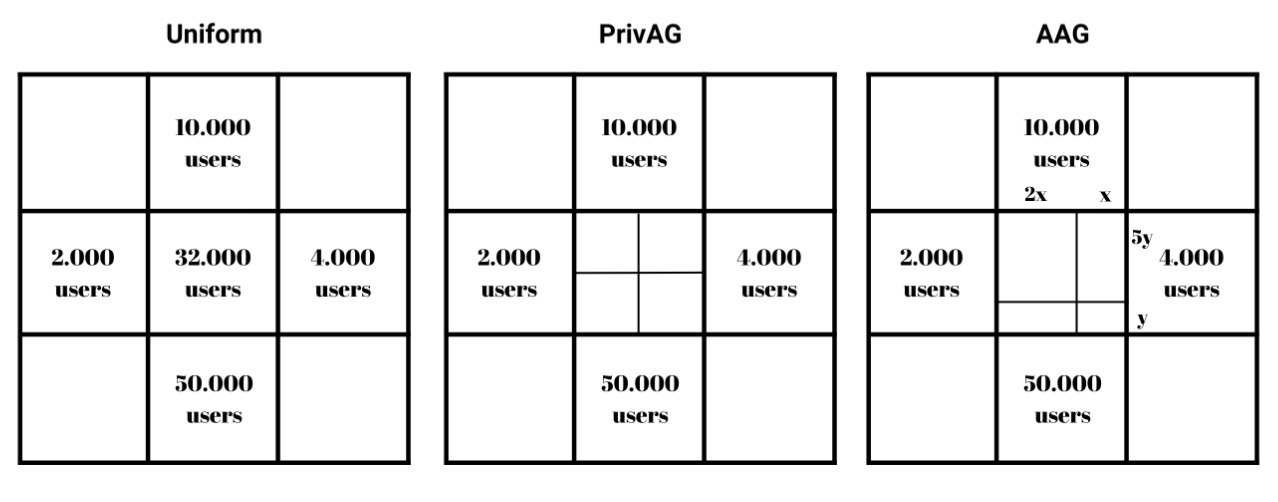}
    \vspace{-10pt}
    \caption{Difference between PrivAG and AAG}
    \label{fig:grids}
    \vspace{-6pt}
\end{figure}

As a key contribution of our paper, we propose the Advanced Adaptive Grid (AAG) approach which advances PrivAG. Consider Figure \ref{fig:grids}, which exemplifies a uniform grid with user densities written inside the cells (on the left), PrivAG (in the middle), and our proposed AAG (on the right). Say that the first phase of PrivAG laid a 3 $\times$ 3 uniform grid $\mathcal{G}_1$ on $\Omega$, and the resulting densities of cells are shown on the left of Figure \ref{fig:grids}. In its second phase, PrivAG iterates through each $C_k \in \mathcal{G}_1$ and decides how to further divide $C_k$. Say that in the current iteration, $C_k$ is the middle cell in Figure \ref{fig:grids}, and it is found that $g_2$ = 2. Then, PrivAG divides the middle cell into 2 $\times$ 2 = 4 equally sized cells as shown in Figure \ref{fig:grids}. Our critical intuition is that the division of this middle cell into equal-sized cells is suboptimal. This is because there are 10.000 users in the upper neighbor whereas 50.000 users in the lower neighbor. Furthermore, there are 2.000 users in the left neighbor whereas 4.000 users in the right neighbor. Based on these neighbor cells' densities, it is likely that the bottom right corner of $C_k$ is denser whereas the upper left corner is more sparse, because the intra-cell distribution is likely to be mimicked by neighbor cells. According to the original intuition behind adaptive grids \cite{qardaji2013differentially,yang2022collecting}, it is desirable to have many small cells in dense areas but few large cells in sparse areas. Following this intuition, instead of dividing $C_k$ evenly (as done in PrivAG), our AAG approach proposes to divide the cell by taking into account the neighbors' densities. Therefore, the vertical division of the cell is done with the ratio 1-to-5 which is proportional to the densities of the upper and lower neighbors (10.000 vs 50.000), whereas the horizontal division is done with the ratio 1-to-2 which is proportional to the left and right neighbors (2.000 vs 4.000). The result is shown in Figure \ref{fig:grids} (on the right). 

We give an algorithmic overview of our proposed AAG approach in Algorithm \ref{alg:AAG}. The first phase of AAG is identical to PrivAG, i.e., it lays a $g_1 \times g_1$ uniform grid $\mathcal{G}_1$ on $\Omega$ and obtains the cell densities $\Phi(C_1)$, $\Phi(C_2)$, ... for all $C_k \in \mathcal{G}_1$ using $\varepsilon$-LDP. The core difference lies in the second phase of AAG. After computing $g_2^k$, instead of dividing cell $C_k$ uniformly, AAG performs the following. First, it calculates the horizontal split location using the densities of the left and right neighbors. That is, denoting the left neighbor of $C_k$ by $C^{L}_k$ and the right neighbor of $C_k$ by $C^{R}_k$, the horizontal split location $hsplit$ is calculated as:
\begin{equation} \label{eq:hsplit}
    hsplit = \frac{\Phi(C^{R}_k)}{\Phi(C^{L}_k) + \Phi(C^{R}_k)} \times (\text{width of } C_k)
\end{equation}
Similarly, to calculate the vertical split location, the densities of the upper and lower neighbors are used. Denoting the upper neighbor of $C_k$ by $C^{U}_k$ and the lower neighbor of $C_k$ by $C^{B}_k$, the vertical split location $vsplit$ is calculated as:
\begin{equation} \label{eq:vsplit}
    vsplit = \frac{\Phi(C^{B}_k)}{\Phi(C^{U}_k) + \Phi(C^{B}_k)} \times (\text{height of } C_k)
\end{equation}
Then, $C_k$ is divided horizontally using $hsplit$ and vertically using $vsplit$. As a result, four subcells of $C_k$ are obtained (as exemplified in the rightmost example in Figure \ref{fig:grids}). If $g_2^k > 2$, then each of the four subcells needs to be further divided. This further division is done uniformly, i.e., uniformly into $(g_2^k-1)/2$ pieces horizontally and $(g_2^k-1)/2$ pieces vertically, to make sure that $C_k$ is divided into $g_2^k \times g_2^k$ subcells overall. 




\begin{algorithm}[!t]
\caption{Algorithmic summary of the AAG approach}
\label{alg:AAG}
\begin{algorithmic}[1]
\State \textbf{Input:} Users $\mathcal{U}$, parameters $\alpha$ and $\sigma$, privacy budget $\varepsilon$
\State \textbf{Output:} Densities of each cell in adaptive grid $\mathcal{G}_{aag}$
\\
\LineComment{\textbf{First phase of AAG}}
\State The first phase of AAG is the same as PrivAG

\\
\LineComment{\textbf{Second phase of AAG}}
\For{each cell $C_k \in \mathcal{G}_1$}
    \State Compute $g_2^k$ according to Equation \ref{g2_calc}
    \State Compute $hsplit$ for $C_k$ according to Equation \ref{eq:hsplit}
    \State Compute $vsplit$ for $C_k$ according to Equation \ref{eq:vsplit}
    \State Divide $C_k$ into four subcells using $hsplit$ and $vsplit$
    \If{$g_2^k > 2$}
    \State Uniformly divide each subcell into $\frac{g_2^k-1}{2}$ pieces horizontally and vertically
    \EndIf
\EndFor
\State Let $\mathcal{G}_{aag}$ denote the resulting grid after the above divisions
\State Call Algorithm \ref{alg:uniform} with $\mathcal{U}_2$, $\mathcal{G}_{aag}$ and $\varepsilon$ to obtain $\Phi(C_1)$, $\Phi(C_2)$, .. for all cells in $\mathcal{G}_{aag}$

\end{algorithmic}
\end{algorithm}


\textbf{Handling edge and corner cells.} When $C_k$ is a cell that is located on one of the edges or corners of $\mathcal{G}_1$, it will lack one or more neighbors. For example, consider the top left cell in Figure \ref{fig:grids}, which is a corner cell. This cell has a lower neighbor and a right neighbor, therefore $\Phi(C^{B}_k)$ and $\Phi(C^{R}_k)$ can be found. However, it does not have an upper neighbor or a left neighbor, therefore $\Phi(C^{U}_k)$ and $\Phi(C^{L}_k)$ are not available. Similarly, if a cell is an edge cell, then it has three neighbors but it lacks one neighbor. For example, a cell that is on the rightmost edge of $\mathcal{G}_1$ has its left, upper, and lower neighbors, but it lacks a right neighbor. For cells that lack one or more neighbors, computing their $hsplit$ and $vsplit$ locations via Equations \ref{eq:hsplit} and \ref{eq:vsplit} using zero densities for the missing neighbors leads to erroneous results. To address this problem, we perform the following. If any of the neighbors of the current cell $C_k$ is missing, then $C_k$ uses its own density $\Phi(C_k)$ in place of the missing neighbor's density. For example, for the top left cell which lacks an upper neighbor and left neighbor, instead of assuming $\Phi(C^{U}_k) = 0$ and $\Phi(C^{L}_k) = 0$, we enforce: $\Phi(C^{U}_k) = \Phi(C_k)$ and $\Phi(C^{L}_k) = \Phi(C_k)$. 

\textbf{Choice of $g_1$ and $g_2^k$.} Recall that the $\alpha$ and $\sigma$ parameters in PrivAG affect the values of $g_1$ and $g_2^k$. As we experimented with PrivAG and AAG, we observed that the recommended values for the $\alpha$ and $\sigma$ parameters in PrivAG yield adaptive grids $\mathcal{G}_{aag}$ with cell counts that are similar to the initial uniform grid $\mathcal{G}_1$. This caused our $\mathcal{G}_{aag}$ to be similar to $\mathcal{G}_1$, which diminishes the benefits of using an adaptive grid. To address this problem, we propose to use different values for the $\alpha$ and $\sigma$ parameters in AAG, leading to different choices of $g_1$ and $g_2^k$. In essence, our choices aim to obtain an increased number of cell divisions in dense regions so that dense regions can be partitioned and represented in more detail, but without causing excessively large $g_2^k$. Specifically, as opposed to the default values of $\alpha$ and $\sigma$ in PrivAG, we use $\alpha$ = 0.25 and $\sigma$ = 0.5 in AAG. The increased value of $\alpha$ in AAG directly increases the chosen $g_1$ and $g_2^k$ values, per Equations \ref{g1_calc} and \ref{g2_calc}. Increased $\sigma$ has two effects. First, it causes the estimated densities $\Phi(C_k)$ used in Equation \ref{g2_calc} to be more accurate, since a larger $\mathcal{U}_1$ is used to estimate them. Thus, the accuracy of $g_2^k$ computations will be increased. Second, since $\sqrt{(1-\sigma)}$ appears in Equation \ref{g2_calc}, it slightly decreases the value of $g_2^k$. This slight decrease is much less pronounced than the increase caused by $\sigma$, and it positively contributes to the calculation of $g_2^k$ by ensuring that the chosen $g_2^k$ will not be excessively large. 

\begin{table}[!t]
\centering
\caption{Number of resulting cells in different grid approaches}
\label{tab:cell_counts}
\begin{tabular}{|c|c|c|c|c|c|}
\hline
\textbf{~Dataset~} & \textbf{~Grid~} & ~$\varepsilon$ = 0.5~ & ~~$\varepsilon$ = 1~~ & ~~$\varepsilon$ = 3~~ & ~~$\varepsilon$ = 5~~ \\ \hline
\multirow{3}{*}{Gowalla} & ~Initial $\mathcal{G}_1$~ & 36 & 81 & 324 & 900 \\
& PrivAG & 42 & 96 & 440  & 1358  \\
   & AAG & 276 & 567 & 2451 & 8745   \\
 \hline
\multirow{3}{*}{Porto} & Initial $\mathcal{G}_1$ & 25 & 49 & 225 & 625   \\
  & PrivAG & 31 & 63 & 256  & 745  \\
  & AAG & 188 & 390 & 1567  & 4503  \\
  \hline
\multirow{3}{*}{~Foursquare~} & Initial $\mathcal{G}_1$ & 16 & 36 & 121 & 361 \\
  & PrivAG & 19 & 45 & 162 & 455  \\
   & AAG & 107 & 235 & 957 & 2794 \\
   \hline
\end{tabular}
\end{table}

In Table \ref{tab:cell_counts}, we provide a subset of our experiment results showing the number of cells in the initial grid $\mathcal{G}_1$, the final grid produced by PrivAG, and the final grid produced by AAG. We repeat this experiment using three datasets: Gowalla, Porto, Foursquare (more details about the datasets are provided in Section \ref{sec:Experiments}), and $\varepsilon$ values ranging between 0.5 and 5. We observe from Table \ref{tab:cell_counts} that especially for lower $\varepsilon$ values such as 0.5 and 1, the number of cells in $\mathcal{G}_1$ is very similar to PrivAG. This supports our motivation that PrivAG's division of dense cells is not sufficiently detailed. In contrast, our choices in AAG ensure that the grids produced by AAG always have a larger number of cells compared to $\mathcal{G}_1$. 

\vspace{-4pt}
\section{Experiments and Discussion} \label{sec:Experiments}

\vspace{-4pt}
\subsection{Experiment Setup and Datasets}

In this section, we experimentally compare the three grid-based decomposition methods (UG, PrivAG, AAG) under varying parameters. We implemented all algorithms and methods in Python. We use three real-world location datasets in our experiments: Gowalla, Porto, and Foursquare. To account for LDP randomness, each experiment is repeated 10 times and the average results are reported.

\textit{Gowalla:} Gowalla was a location-based social networking site where users shared their locations via check-ins. From the full dataset \cite{cho2011friendship}, we extracted check-ins made in the United States, between longitudes -124.26 and -71.87 and latitudes 25.45 and 47.44. Consequently, we have 3,451,190 remaining locations.

\textit{Porto:} The Porto dataset contains trips of 442 taxis driving in the city of Porto. It was released as part of the Taxi Service Prediction Competition in ECML-PKDD \cite{moreira2013predicting}. The original dataset contains full taxi trips, i.e., trajectories with multiple location readings per trip. We pre-processed it by keeping only one randomly sampled location from each trip and treated them as the current locations of users $\mathcal{U}$. We only used the locations between longitudes -8.691294 and -8.552009 and latitudes 41.138351 and 41.185935, corresponding to the city of Porto. This resulted in 1,620,157 remaining locations.

\textit{Foursquare:} The Foursquare dataset contains location check-ins of social media users in Tokyo, between April 2012 and February 2013 \cite{yang2014modeling}. We used this dataset without pre-processing. In total, the dataset contains 573,703 locations. 

\vspace{-4pt}
\subsection{Utility Metric} \label{sec:Utility}

Following previous works, we use spatial density queries for utility measurement \cite{alptekin2023building,gursoy2018utility,qardaji2013differentially,yang2022collecting}. A spatial density query $q$ with geospatial area denoted by $A(q)$ is a query of the form: ``How many users are located within $A(q)$''? Let $ans_{q}$ denote the ground truth answer of $q$, i.e., the answer that would be computed if all users' locations were known exactly, without privacy constraints. Using a grid $\mathcal{G}$ and estimated cell densities $\Phi(C_k)$ for $C_k \in \mathcal{G}$, the noisy answer of $q$ (denoted by $ans'_{q}$) can be calculated as follows. First, $ans'_{q}$ is initialized as 0. Then, each cell $C_k \in \mathcal{G}$ is visited one by one:
\begin{enumerate}
\vspace{-3pt}
    \item Denoting by $A(C_k)$ the geospatial area of $C_k$, if $A(C_k)$ is disjoint from $A(q)$, then $C_k$ is ignored.
    \item If $A(C_k)$ is fully contained within $A(q)$, then $ans'_{q}$ is increased by $\Phi(C_k)$. 
    \item If $A(C_k)$ partially intersects with $A(q)$, then $ans'_{q}$ is increased proportional to the intersection amount, i.e.:
    \begin{equation}
    ans'_q = ans'_q + \Phi(C_k) \times \frac{||A(C_k) \cap A(q)||}{||A(C_k)||}
\end{equation}
\end{enumerate}

\textbf{Average Query Error (AQE):} We generate $\gamma$ = 500 number of random queries $q_1, q_2, ...$ with different $A(q_i)$ and compute their $ans_{q_i}$ and $ans'_{q_i}$. Then, we measure the average error between $ans_{q_i}$ and $ans'_{q_i}$ using the AQE metric:
\begin{equation} \label{eq:AQE}
    AQE = \frac{1}{\gamma} * \sum\limits_{i = 1}^{\gamma} \frac{|ans_{q_i} - ans'_{q_i}|}{max \{ans_{q_i}, b\} }
\end{equation} 
Here, $b$ denotes a bound to mitigate the dominating effect of queries with extremely high selectivities (extremely low $ans_{q_i}$) \cite{gursoy2018differentially,gursoy2018utility}. We set the value of $b$ as: $b = 2\% \times |\mathcal{U}|$. 

\vspace{-3pt}
\subsection{Uniform Grid Experiments} \label{sec:UGexperiments}

\begin{figure*}[!t]
\centering
     \includegraphics[width=.30\textwidth]{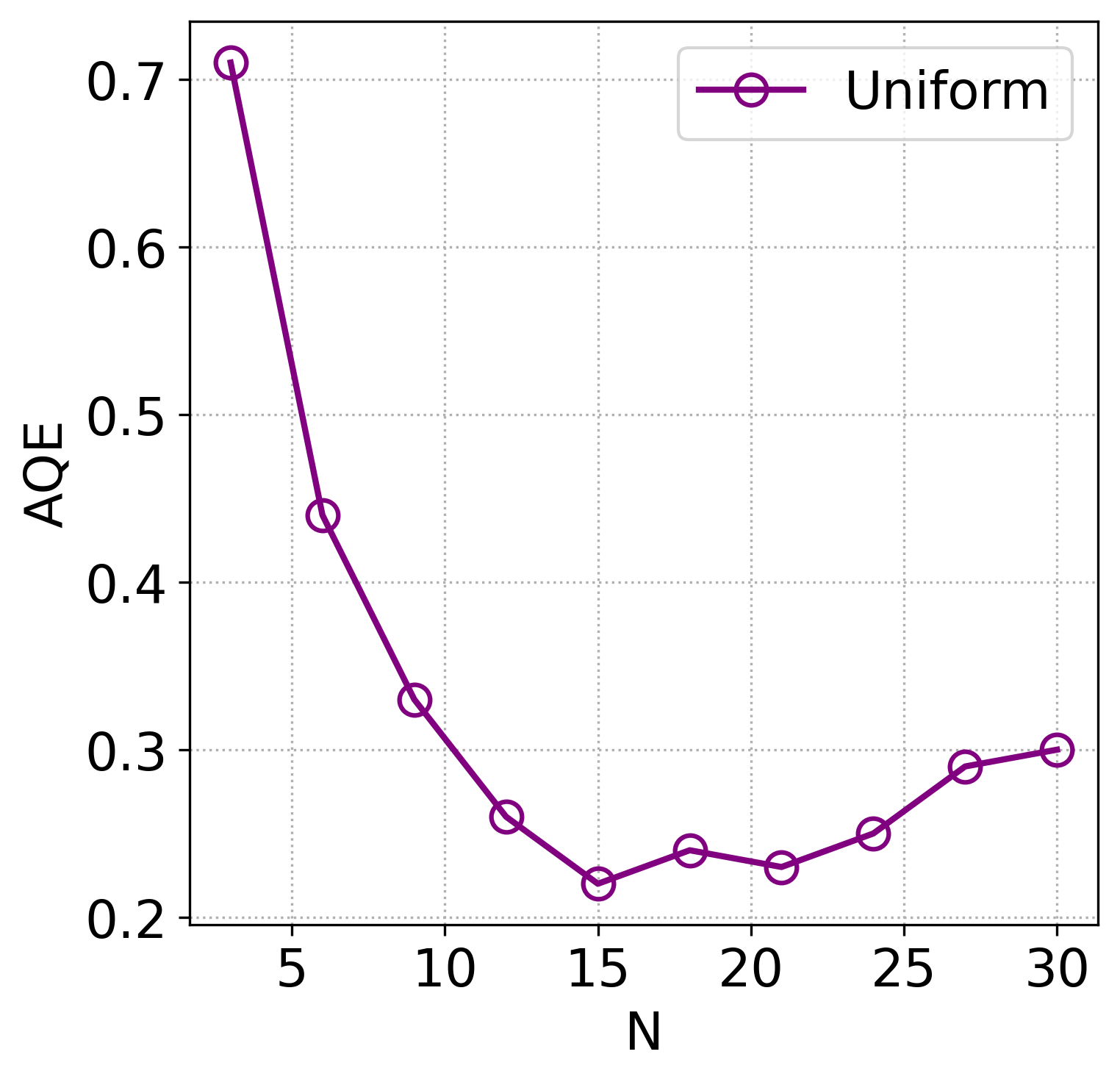}
     \hspace{2pt}
     \includegraphics[width=.30\textwidth]{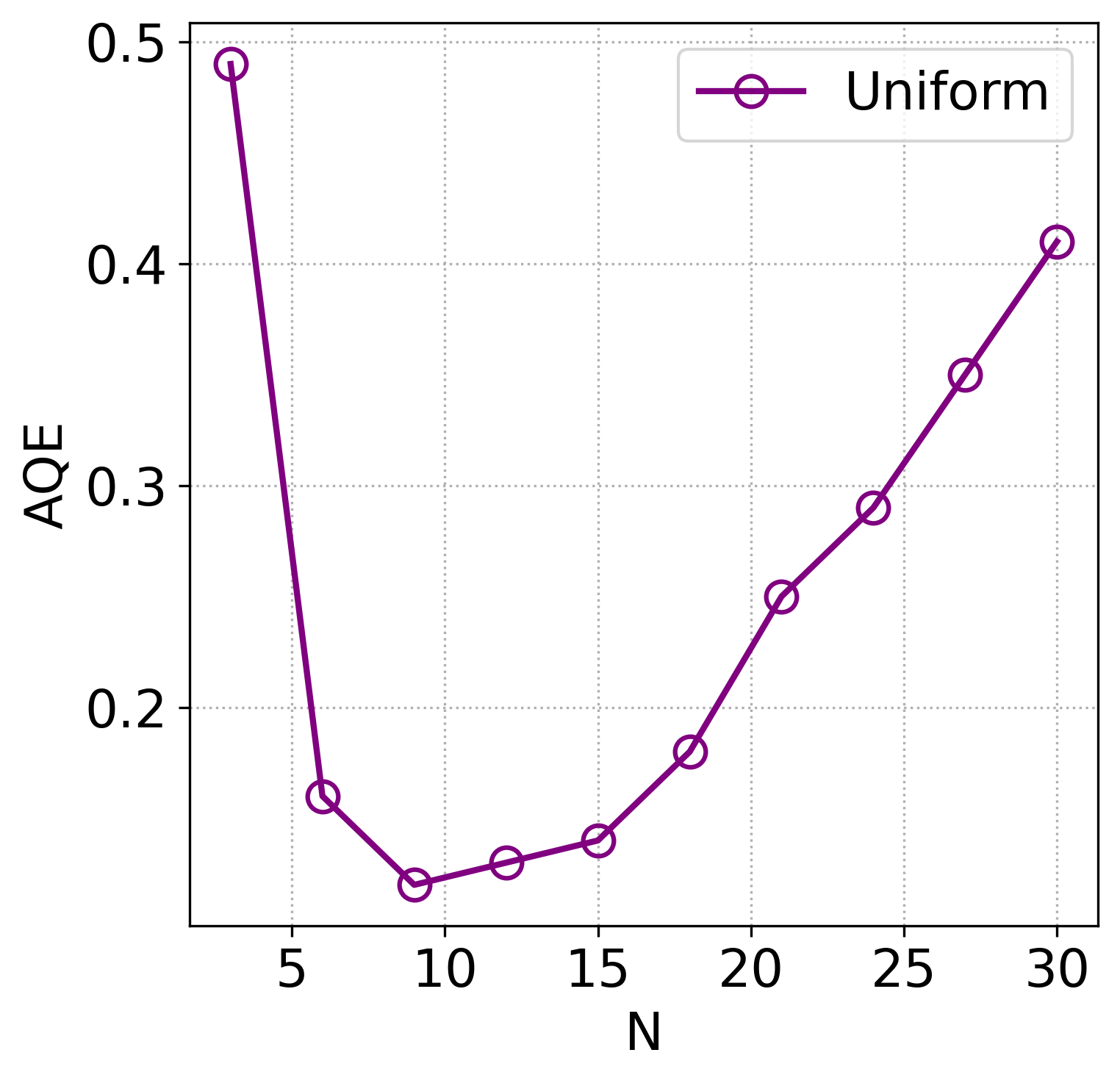}
     \hspace{2pt}
     \includegraphics[width=.31\textwidth]{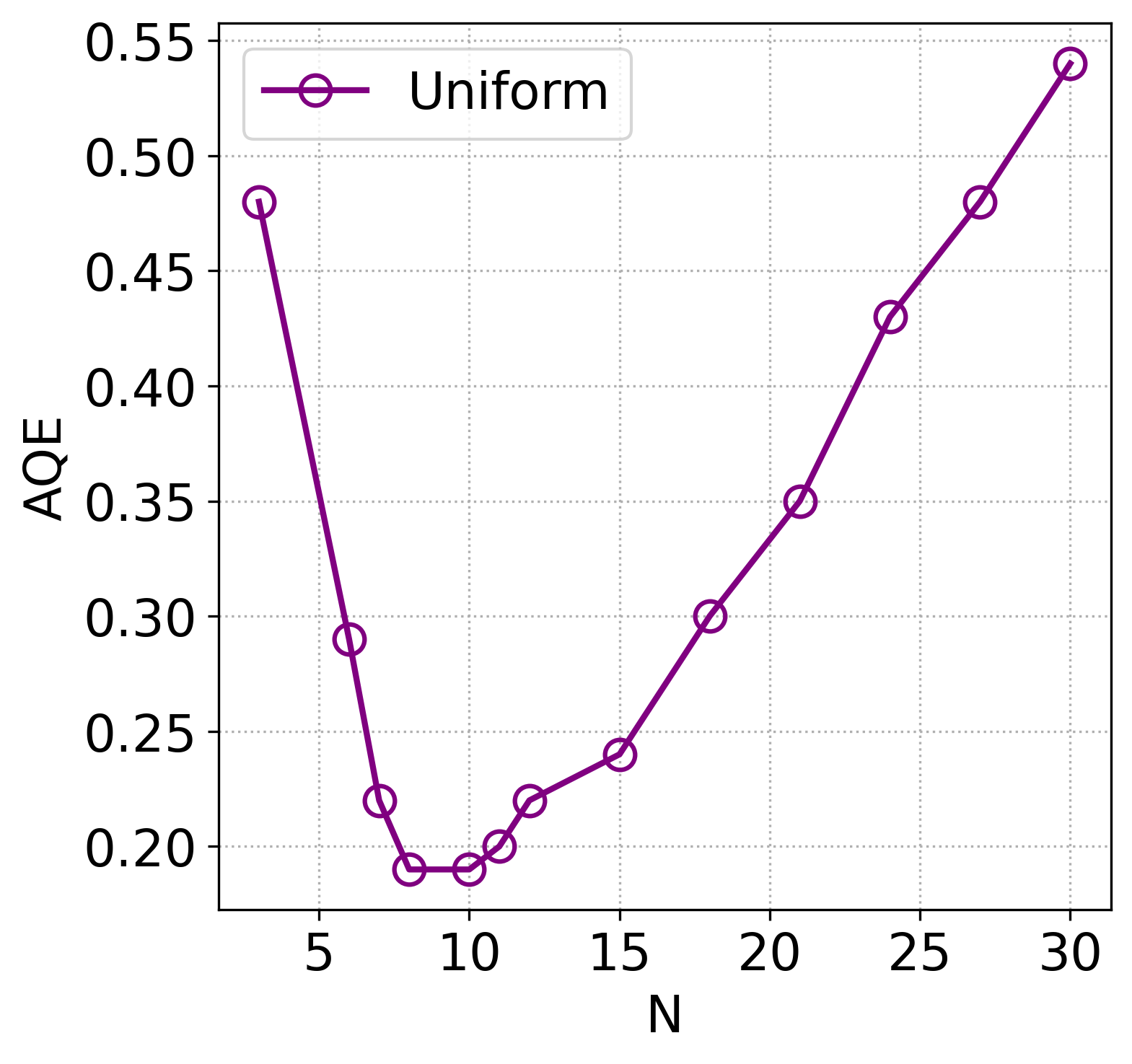}
     \vspace{-8pt}
    \caption{AQEs of $N \times N$ uniform grids $\mathcal{G}_{uni}$ with varying $N$, fixed $\varepsilon$ = 1. From left to right: Gowalla dataset, Porto dataset, Foursquare dataset.}
    \vspace{-4pt}
    \label{fig:uniform}
\end{figure*}

To measure the utility impact of the number of cells on the Uniform Grid (UG), we generate $N \times N$ uniform grids with various $N$ and measure their AQEs. The results are shown in Figure \ref{fig:uniform}. It can be observed from Figure \ref{fig:uniform} that AQEs are highly dependent on the value of $N$. If $N$ is badly chosen (e.g., too small or too large), then AQEs can be 3-4 times higher compared to the case when $N$ is near-optimal. Furthermore, results on all three datasets follow a U-shaped curve, e.g., $N$ between 15-20 seems to be optimal for the Gowalla dataset, $N$ = 10 seems to be optimal for the Porto dataset, and $N$ between 7-12 seems to be optimal for the Foursquare dataset. The fact that different $N$ values are best for different datasets makes it difficult to determine a good value of $N$ prior to data collection in practice. Overall, we can conclude that the UG approach is highly sensitive to the choice of $N$, and the optimal choice of $N$ varies from one dataset to another. In the rest of the experiments, we use the best results (i.e., optimized choice of $N$) when comparing UG with other approaches. 

\vspace{-4pt}
\subsection{Comparison of Grid Approaches}

In this section, we compare the three grid approaches (UG, PrivAG, AAG) using different-sized random queries. To do so, we enforce that the random queries we generate for calculating AQE have size: $A(q) = \rho \times \Omega$, where $\rho \in (0,1]$ is the query size parameter. In Table \ref{tab:smallRho}, we fix $\varepsilon$ = 1 and vary the value of $\rho$ between 0.005\% and 0.5\%. Note that these are relatively low values of $\rho$, i.e., the generated queries are small. We use the \textbf{bold} notation in Table \ref{tab:smallRho} when comparing the two adaptive grid approaches (PrivAG vs AAG), i.e., the one that yields lower AQE is written in bold. This helps to demonstrate the improvement of AAG compared to PrivAG. We use the grey cell background to denote the best-performing approach when comparing all three (UG vs PrivAG vs AAG). 

\begin{table}[!t]
\centering
\caption{AQEs with varying query sizes $\rho$, fixed $\varepsilon$ = 1.}
\label{tab:smallRho}
\begin{tabular}{|c|c|c|c|c|c|c|}
\hline
\textbf{~Dataset~} & \textbf{~Method~} & ~$\rho$ = 0.005\%~ & ~$\rho$ = 0.01\%~ & ~$\rho$ = 0.05\%~ & ~$\rho$ = 0.1\%~ & ~$\rho$ = 0.5\%~ \\ \hline
\multirow{3}{*}{Gowalla} & ~UG~ &  0.0034 & 0.0067 & 0.0279 & 0.0485 & \cellcolor{lightgray}0.120 \\
& PrivAG & 0.0039 & 0.0077 & 0.0374 & 0.0728 & 0.305 \\
   & AAG & \cellcolor{lightgray}\textbf{0.0023} & \cellcolor{lightgray}\textbf{0.0051} & \cellcolor{lightgray}\textbf{0.0236} & \cellcolor{lightgray}\textbf{0.0460} & \textbf{0.185} \\
 \hline
\multirow{3}{*}{Porto} & UG & 0.0028 & \cellcolor{lightgray}0.0045 & \cellcolor{lightgray}0.0180 & \cellcolor{lightgray}0.0321 & \cellcolor{lightgray}0.082  \\
  & PrivAG & 0.0034 & 0.0056 & 0.0283 & 0.0540 & 0.205 \\
  & AAG & \cellcolor{lightgray}\textbf{0.0025} & \textbf{0.0045} & \textbf{0.0247} & \textbf{0.0501} & \textbf{0.195} \\
  \hline
\multirow{3}{*}{Foursquare} & UG & 0.0032 & 0.0054 & 0.0243 & \cellcolor{lightgray}0.0416 & \cellcolor{lightgray}0.126 \\
  & PrivAG & 0.0036 & 0.0062 & 0.0291 & 0.0547 & 0.203  \\
   & AAG & \cellcolor{lightgray}\textbf{0.0025} & \cellcolor{lightgray}\textbf{0.0043} & \cellcolor{lightgray}\textbf{0.0234} & \textbf{0.0450} & \textbf{0.177} \\
   \hline
\end{tabular}
\vspace{-4pt}
\end{table}

The results in Table \ref{tab:smallRho} show that AAG achieves lower error compared to PrivAG in all settings. In addition, AAG also achieves lower error compared to UG in the majority of settings. However, as $\rho$ increases, UG starts performing better than AAG. This implies that for fine-grained density modeling and small-sized queries, AAG is the best approach. This is an intuitive result, considering that AAG excels in dividing dense areas in a detailed fashion. Yet, for more coarse (high-level) density statistics, UG can perform better. Another observation we make from Table \ref{tab:smallRho} is that when $\rho$ increases, the errors also increase. This is because increasing $\rho$ causes the intersection between $A(q_i)$ and various cells to increase, therefore $ans_{q_i}$ and $ans'_{q_i}$ become larger. Hence, overall noise amount increases as well, and among the two factors in the denominator of Equation \ref{eq:AQE}, $ans_{q_i}$ starts to dominate rather than $b$. Consequently, higher AQEs are obtained.

In Table \ref{tab:largeRho}, we perform the same experiment but with much larger $\rho$ values, ranging between 2\% and 10\%. For such large $\rho$ values, we observe that UG achieves lower AQE compared to PrivAG and AAG in all cases. It should be noted that these results use optimized $N$ values for UG, hence they do not account for potentially suboptimal choices of $N$. On the other hand, comparing PrivAG versus AAG, we once again observe that AAG beats PrivAG. 


\begin{table}[!t]
\centering
\caption{AQEs with varying query sizes $\rho$, fixed $\varepsilon$ = 1. Much larger $\rho$ values are used in this table compared to Table \ref{tab:smallRho}.}
\label{tab:largeRho}
\begin{tabular}{|c|c|c|c|c|c|c|}
\hline
\textbf{~Dataset~} & \textbf{~Method~} & ~~$\rho$ = 2\%~~ & ~~$\rho$ = 4\%~~ & ~~$\rho$ = 6\%~~ & ~~$\rho$ = 8\%~~ & ~~$\rho$ = 10\%~~ \\ \hline
\multirow{3}{*}{Gowalla} & ~UG~ & \cellcolor{lightgray}0.20 & \cellcolor{lightgray}0.21 & \cellcolor{lightgray}0.24 & \cellcolor{lightgray}0.22 & \cellcolor{lightgray}0.21 \\
& PrivAG & 0.83 & 1.15 & 1.32 & 1.29 & 1.25 \\
   & AAG & \textbf{0.52} & \textbf{0.71} & \textbf{0.73} & \textbf{0.73} & \textbf{0.72} \\
 \hline
\multirow{3}{*}{Porto} & UG & \cellcolor{lightgray}0.12 & \cellcolor{lightgray}0.12 & \cellcolor{lightgray}0.11 & \cellcolor{lightgray}0.12 & \cellcolor{lightgray}0.09 \\
  & PrivAG & 0.54 & 0.76 & 0.85 & 0.91 & 0.91 \\
  & AAG & \textbf{0.38} & \textbf{0.49} & \textbf{0.60} & \textbf{0.61} & \textbf{0.65} \\
  \hline
\multirow{3}{*}{~Foursquare~} & UG & \cellcolor{lightgray}0.17 & \cellcolor{lightgray}0.20 & \cellcolor{lightgray}0.19 & \cellcolor{lightgray}0.22 & \cellcolor{lightgray}0.21 \\
  & PrivAG & 0.48 & 0.63 & 0.68 & 0.70 & 0.71 \\
   & AAG & \textbf{0.40} & \textbf{0.56} & \textbf{0.64} & \textbf{0.70} & \textbf{0.71} \\
   \hline
\end{tabular}
\end{table}

\vspace{-4pt}
\subsection{Impact of the Privacy Budget $\varepsilon$} \label{sec:ImpactEpsilon}

In this section, we keep the query sizes $\rho$ fixed and vary the privacy budgets $\varepsilon$ between 0.5 and 5. Table \ref{tab:eps001} provides the results with $\rho$ = 0.01\% and Table \ref{tab:eps4} provides the results with $\rho$ = 4\%. In both tables, we use the same bold and grey color highlight strategies that we used in the previous section. According to the results in Table \ref{tab:eps001}, when $\rho$ =  0.01\%, AAG is the best approach. It yields the lowest AQEs across all $\varepsilon$. On the other hand, when $\rho$ = 4\%, UG becomes the best approach as shown in Table \ref{tab:eps4}. This is parallel to the results reported in the previous section. When $\rho$ = 4\%, although AAG consistently beats PrivAG, it cannot reach UG's low AQE values.

\begin{table}[!t]
\centering
\caption{AQEs with varying privacy budgets $\varepsilon$, fixed $\rho$ = 0.01\%.}
\label{tab:eps001}
\begin{tabular}{|c|c|c|c|c|c|}
\hline
\textbf{~Dataset~} & \textbf{~Method~} & ~~$\varepsilon$ = 0.5~~ & ~~$\varepsilon$ = 1~~ & ~~$\varepsilon$ = 3~~ & ~~$\varepsilon$ = 5~~  \\ \hline
\multirow{3}{*}{Gowalla} & ~UG~ & 0.0070 & 0.0066 & 0.0064 & 0.0056 \\
& PrivAG & 0.0075 & 0.0077 & 0.0072 & 0.0062 \\
   & AAG & \cellcolor{lightgray}\textbf{0.0047} & \cellcolor{lightgray}\textbf{0.0051} & \cellcolor{lightgray}\textbf{0.0049} & \cellcolor{lightgray}\textbf{0.0041} \\
 \hline
\multirow{3}{*}{Porto} & UG & 0.0048 & 0.0045 & \cellcolor{lightgray}0.0045 & \cellcolor{lightgray}0.0045 \\
  & PrivAG & 0.0058 & 0.0056 & 0.0059 & 0.0060 \\
  & AAG & \cellcolor{lightgray}\textbf{0.0048} & \cellcolor{lightgray}\textbf{0.0045} & \textbf{0.0049} & \textbf{0.0049} \\
  \hline
\multirow{3}{*}{~Foursquare~} & UG & 0.0055 & 0.0054 & 0.0051 & 0.0048 \\
  & PrivAG & 0.0060 & 0.0062 & 0.0061 & 0.0069 \\
   & AAG & \cellcolor{lightgray}\textbf{0.0044} & \cellcolor{lightgray}\textbf{0.0043} & \cellcolor{lightgray}\textbf{0.0040} & \cellcolor{lightgray}\textbf{0.0047} \\
   \hline
\end{tabular}
\vspace{-4pt}
\end{table}

\begin{table}[!t]
\centering
\caption{AQEs with varying privacy budgets $\varepsilon$, fixed $\rho$ = 4\%.}
\label{tab:eps4}
\begin{tabular}{|c|c|c|c|c|c|}
\hline
\textbf{~Dataset~} & \textbf{~Method~} & ~~$\varepsilon$ = 0.5~~ & ~~$\varepsilon$ = 1~~ & ~~$\varepsilon$ = 3~~ & ~~$\varepsilon$ = 5~~  \\ \hline
\multirow{3}{*}{Gowalla} & ~UG~ & \cellcolor{lightgray}0.28 & \cellcolor{lightgray}0.23 & \cellcolor{lightgray}0.17 & \cellcolor{lightgray}0.19 \\
& PrivAG & 1.09 & 1.15 & 1.08 & 0.98 \\
   & AAG & \textbf{0.68} & \textbf{0.71} & \textbf{0.68} & \textbf{0.76} \\
 \hline
\multirow{3}{*}{Porto} & UG & \cellcolor{lightgray}0.16 & \cellcolor{lightgray}0.12 & \cellcolor{lightgray}0.10 & \cellcolor{lightgray}0.09 \\
  & PrivAG & 0.77 & 0.76 & 0.71 & 0.71 \\
  & AAG & \textbf{0.49} & \textbf{0.49} & \textbf{0.62} & \textbf{0.63} \\
  \hline
\multirow{3}{*}{~Foursquare~} & UG & \cellcolor{lightgray}0.26 & \cellcolor{lightgray}0.19 & \cellcolor{lightgray}0.16 & \cellcolor{lightgray}0.16 \\
  & PrivAG & 0.69 & 0.63 & 0.75 & 0.69 \\
   & AAG & \textbf{0.61} & \textbf{0.56} & \textbf{0.55} & \textbf{0.51} \\
   \hline
\end{tabular}
\end{table}

In both tables, we observe that as $\varepsilon$ increases, AQEs of UG decrease. This is an intuitive result since higher $\varepsilon$ means less perturbation caused by LDP, therefore results are more accurate. On the other hand, although this trend holds for UG, it does not always hold for PrivAG and AAG. For example, despite increasing $\varepsilon$, there are cases in PrivAG and AAG in which AQE values increase. The reason is that, as shown in Equations \ref{g1_calc} and \ref{g2_calc}, $\varepsilon$ is used in the choice of $g_1$ and $g_2^k$. Hence, changing $\varepsilon$ also changes the grid structures in PrivAG and AAG. These structural changes may affect $ans'_q$ positively or negatively, and LDP perturbation is no longer the only factor in the accuracy of $ans'_q$. Hence, we do not see a consistent trend between $\varepsilon$ and AQE in PrivAG and AAG. On the other hand, this observation shows that if the choices of $g_1$ and $g_2^k$ are made in a more optimized fashion, especially in high $\varepsilon$ regions, there is potential to improve utility. This can be a good avenue for future work. 

Combining all experiment results, we arrive at the following take-away messages: (i) AAG is preferable to PrivAG across all $\varepsilon$ and $\rho$, (ii) AAG is the best approach when $\rho$ is small, e.g., for computing answers to small queries or for detailed statistics, and (iii) UG is the best approach when $\rho$ is large, e.g., for computing answers to large queries or for coarse statistics.

\vspace{-4pt}
\section{Related Work} \label{sec:RelatedWork}

The rising popularity of LDP in recent years has led to growing interest in its application to spatial data, given the widespread use of location-based services (LBS) and the importance of protecting location privacy. Chen et al.~\cite{chen2016private} proposed a variant of LDP called personalized LDP (PLDP), and developed methods for spatial data aggregation under PLDP. Wang et al.~\cite{wang2021lsrr} developed L-SRR to enable LBS tasks while satisfying LDP. Hong et al.~\cite{hong2022collecting} studied the problem of reducing the expected error of each perturbed location when collecting location data with LDP. Kim et al.~\cite{kim2018application} and Navidan et al.~\cite{navidan2022hide} applied LDP to indoor positioning. Cunningham et al.~\cite{cunningham2021real} proposed a mechanism for trajectory sharing under LDP using n-grams, and Zhang et al.~\cite{zhang2023trajectory} proposed a mechanism for trajectory perturbation using points' adjacent direction information to improve utility. Du et al.~\cite{du2023ldptrace} proposed LDPTrace for synthesizing realistic location trajectories under LDP, which utilizes a uniform grid for discretization.

Hu et al.~\cite{hu2023continuous} studied the problem of continuously releasing location statistics from streaming data while protecting privacy using $w$-event level LDP. Alptekin and Gursoy \cite{alptekin2023building} studied the problem of building quadtrees for spatial data under LDP. Yao et al.~\cite{yao2024privacy} proposed a framework for collecting users' locations under LDP for Unmanned Aerial Vehicles (UAVs), utilizing a quadtree-based approach. Most closely related to our work is the work of Yang et al.~\cite{yang2022collecting}, who studied the problem of collecting location trajectories under LDP. For this problem, they developed an adaptive grid approach called PrivAG. We experimentally show that our AAG improves PrivAG.


\vspace{-6pt}
\section{Conclusion} \label{sec:Conclusion}
\vspace{-4pt}

In this paper, we studied three grid-based decomposition approaches under LDP: Uniform Grid (UG), PrivAG, and AAG. Our proposed AAG approach advances the state-of-the-art adaptive grid approach (PrivAG) by performing cell divisions according to neighboring cells' densities. We experimentally compared UG, PrivAG, and AAG using three datasets and multiple parameter values ($\varepsilon$ and $\rho$). We observed that AAG always beats PrivAG, and it also beats UG when $\rho$ is small. However, when $\rho$ is large, UG with a near-optimal choice of grid size becomes better than AAG. In future work, we plan to compare these three grid approaches against tree-based decompositions. Furthermore, we will investigate methods to improve PrivAG and AAG's utility especially in high $\varepsilon$ regimes.

\vspace{-6pt}
\section*{Acknowledgments}
\vspace{-6pt}

This study was supported by Scientific and Technological Research Council of Türkiye (TUBITAK) under Grant Number 121E303. The authors thank TUBITAK for their support.

\vspace{-6pt}

\bibliographystyle{splncs04}
\bibliography{references}

\begin{thebibliography}{10}
\providecommand{\url}[1]{\texttt{#1}}
\providecommand{\urlprefix}{URL }
\providecommand{\doi}[1]{https://doi.org/#1}

\bibitem{alptekin2023building}
Alptekin, E., Gursoy, M.E.: Building quadtrees for spatial data under local differential privacy. In: IFIP Annual Conference on Data and Applications Security and Privacy. pp. 22--39. Springer (2023)

\bibitem{chen2016private}
Chen, R., Li, H., Qin, A., Kasiviswanathan, S.P., Jin, H.: Private spatial data aggregation in the local setting. In: 2016 IEEE 32nd International Conference on Data Engineering (ICDE). pp. 289--300. IEEE (2016)

\bibitem{cho2011friendship}
Cho, E., Myers, S.A., Leskovec, J.: Friendship and mobility: user movement in location-based social networks. In: 17th ACM SIGKDD International Conference on Knowledge Discovery and Data Mining. pp. 1082--1090 (2011)

\bibitem{cormode2018privacy}
Cormode, G., Jha, S., Kulkarni, T., Li, N., Srivastava, D., Wang, T.: Privacy at scale: Local differential privacy in practice. In: Proceedings of the 2018 International Conference on Management of Data. pp. 1655--1658. ACM (2018)

\bibitem{cormode2021frequency}
Cormode, G., Maddock, S., Maple, C.: Frequency estimation under local differential privacy. Proceedings of the VLDB Endowment  \textbf{14}(11),  2046--2058 (2021)

\bibitem{cunningham2021real}
Cunningham, T., Cormode, G., Ferhatosmanoglu, H., Srivastava, D.: Real-world trajectory sharing with local differential privacy. Proceedings of the VLDB Endowment  \textbf{14}(11),  2283--2295 (2021)

\bibitem{ding2017collecting}
Ding, B., Kulkarni, J., Yekhanin, S.: Collecting telemetry data privately. In: Advances in Neural Information Processing Systems. pp. 3571--3580 (2017)

\bibitem{du2023ldptrace}
Du, Y., Hu, Y., Zhang, Z., Fang, Z., Chen, L., Zheng, B., Gao, Y.: Ldptrace: Locally differentially private trajectory synthesis. Proceedings of the VLDB Endowment  \textbf{16}(8),  1897--1909 (2023)

\bibitem{erlingsson2014rappor}
Erlingsson, {\'U}., Pihur, V., Korolova, A.: Rappor: Randomized aggregatable privacy-preserving ordinal response. In: Proceedings of the 2014 ACM SIGSAC Conference on Computer and Communications Security. pp. 1054--1067. ACM (2014)

\bibitem{gursoy2022adversarial}
Gursoy, M.E., Liu, L., Chow, K.H., Truex, S., Wei, W.: An adversarial approach to protocol analysis and selection in local differential privacy. IEEE Transactions on Information Forensics and Security  \textbf{17},  1785--1799 (2022)

\bibitem{gursoy2018differentially}
Gursoy, M.E., Liu, L., Truex, S., Yu, L.: Differentially private and utility preserving publication of trajectory data. IEEE Transactions on Mobile Computing  \textbf{18}(10),  2315--2329 (2018)

\bibitem{gursoy2018utility}
Gursoy, M.E., Liu, L., Truex, S., Yu, L., Wei, W.: Utility-aware synthesis of differentially private and attack-resilient location traces. In: Proceedings of the 2018 ACM SIGSAC Conference on Computer and Communications Security. pp. 196--211 (2018)

\bibitem{hong2022collecting}
Hong, D., Jung, W., Shim, K.: Collecting geospatial data under local differential privacy with improving frequency estimation. IEEE Transactions on Knowledge and Data Engineering  (2022)

\bibitem{hu2023continuous}
Hu, R., Li, H., Li, J., Wang, Z., Wang, B.: Continuous release of temporal correlation location statistics with local differential privacy. Multimedia Tools and Applications pp. 1--19 (2023)

\bibitem{kim2018application}
Kim, J.W., Kim, D.H., Jang, B.: Application of local differential privacy to collection of indoor positioning data. IEEE Access  \textbf{6},  4276--4286 (2018)

\bibitem{moreira2013predicting}
Moreira-Matias, L., Gama, J., Ferreira, M., Mendes-Moreira, J., Damas, L.: Predicting taxi--passenger demand using streaming data. IEEE Transactions on Intelligent Transportation Systems  \textbf{14}(3),  1393--1402 (2013)

\bibitem{navidan2022hide}
Navidan, H., Moghtadaiee, V., Nazaran, N., Alishahi, M.: Hide me behind the noise: Local differential privacy for indoor location privacy. In: 2022 IEEE European Symposium on Security and Privacy Workshops. pp. 514--523. IEEE (2022)

\bibitem{qardaji2013differentially}
Qardaji, W., Yang, W., Li, N.: Differentially private grids for geospatial data. In: 2013 IEEE 29th International Conference on Data Engineering (ICDE). pp. 757--768. IEEE (2013)

\bibitem{wang2021lsrr}
Wang, H., Hong, H., Xiong, L., Qin, Z., Hong, Y.: L-srr: Local differential privacy for location-based services with staircase randomized response. In: Proceedings of the 2022 ACM SIGSAC Conference on Computer and Communications Security. p. 2809–2823. Association for Computing Machinery, New York, NY, USA (2022)

\bibitem{wang2017locally}
Wang, T., Blocki, J., Li, N., Jha, S.: Locally differentially private protocols for frequency estimation. In: Proc. of the 26th USENIX Security Symposium. pp. 729--745 (2017)

\bibitem{wang2018locally}
Wang, T., Li, N., Jha, S.: Locally differentially private frequent itemset mining. In: IEEE Symposium on Security and Privacy (SP). IEEE (2018)

\bibitem{yang2014modeling}
Yang, D., Zhang, D., Zheng, V.W., Yu, Z.: Modeling user activity preference by leveraging user spatial temporal characteristics in lbsns. IEEE Transactions on Systems, Man, and Cybernetics: Systems  \textbf{45}(1),  129--142 (2014)

\bibitem{yang2022collecting}
Yang, J., Cheng, X., Su, S., Sun, H., Chen, C.: Collecting individual trajectories under local differential privacy. In: 2022 23rd IEEE International Conference on Mobile Data Management (MDM). pp. 99--108. IEEE (2022)

\bibitem{yao2024privacy}
Yao, A., Pal, S., Li, X., Zhang, Z., Dong, C., Jiang, F., Liu, X.: A privacy-preserving location data collection framework for intelligent systems in edge computing. Ad Hoc Networks  \textbf{161},  103532 (2024)

\bibitem{zhang2023trajectory}
Zhang, Y., Ye, Q., Chen, R., Hu, H., Han, Q.: Trajectory data collection with local differential privacy. Proceedings of the VLDB Endowment  \textbf{16}(10),  2591--2604 (2023)

\end{thebibliography}

\end{document}